\RequirePackage{fix-cm}
\documentclass[smallextended]{svjour3}       
\smartqed  
\usepackage{graphicx}
\begin{document}
\graphicspath{{images/},{../../../echo/plots/},{../../../echo/plots/PCA/},{../../../echo/templates/west/plots/}}

\title{The contribution of the major planet search surveys to EChO target selection}

\author{Giuseppina Micela \and 
	G\'asp\'ar \'A.~Bakos\footnote{Sloan Fellow, Packard Fellow} \and Mercedes Lopez-Morales \and Pierre F. L. Maxted \and Isabella Pagano \and Alessandro Sozzetti \and Peter J. Wheatley}


\institute{
G. Micela \at
INAF - Osservatorio Astronomico di Palermo,\\
Piazza del Parlamento 1, 90134 Palermo, Italy\\
\email giusi@astropa.inaf.it
\and
G. Bakos\ \at
Princeton University\\
4 Ivy lane, Princeton, NJ08544, USA
\and
M.  Lopez Morales \at 
Harvard-Smithsonian Center for Astrophysics, \\
60 Garden Street, Cambridge, MA 02138, USA
\and
P. F. L. Maxted \at
Astrophysics Group, Keele University, \\
Keele, Staffordshire ST5 5BG, UK
\and
I. Pagano \at
              INAF - Osservatorio Astrofisico di Catania,\\
              via S. Sofia, 78, 95123 Catania, Italy
\and 
A. Sozzetti \at
INAF - Osservatorio Astrofisico di Torino,\\
Via Osservatorio 20, I-10025 Pino Torinese, Italy
\and
P. J. Wheatley   \at
        Department of Physics, \\
        University of Warwick, Coventry CV4 7AL, UK
}

\date{Received: date / Accepted: date}

\maketitle

\begin{abstract}
The EChO core science will be based on a three tier survey, each with increasing sensitivity, in order to study the population of exo-planets from super-Earths to Jupiter-like planets, in the very hot to temperate zones (temperatures of 300 K - 3000 K) of F to M-type host stars. To achieve a meaningful outcome an accurate selection of the target sample is needed. In this paper we analyse the targets, suitable for EChO observations, expected to result from a sample of  present and forthcoming detection surveys.

Exoplanets currently known are already sufficient to provide a
large and diverse sample. However we expect the results from these surveys to increase the sample of smaller planets that will allow us to optimize the EChO sample selection.

\end{abstract}

\section{Introduction}\label{sec1}

Target selection is a key aspect of EChO. The choice of the targets will determine the planetary parameter space we will explore. The scientific outcome of the mission will depend on the observed sample.

While it is not required to select the sample ten years in advance, a good plan to select the best sample before lunch is needed. In the present phase we are defining the primary physical planetary parameters that span the ÒdiversityÓ of planet population. These include: 

\begin{itemize}	
\item
Stellar temperature, metallicity, and age, 
\item
Planet temperature, mass and density.
\end{itemize}

A sub-space of this parameter space will be explored by EChO. The mission is designed to  fill such sub-space, and we will need enough of these targets in early 2020's from now. The exo-planetary population known today allows us to fill most of the EChO objectives, however we expect that new discoveries will come out in the next years enabling us to optimize the EChO sample. Several surveys both from ground and from space will provide targets with the necessary characteristics to meet the objectives of the mission.

The following sections describe the most important surveys from which we expect targets that will be observed by EChO. The list is not exhaustive. We note that some classes of objects (such as for example the Hot Jupiters) are abundantly represented in the samples already known today.  We expect the results from these surveys to increase the sample of smaller planets.

\section{WASP and Super-WASP}\label{sec2}

The WASP survey uses two instruments located at the Observatorio del Roque de los Muchachos in Spain and Sutherland in South Africa to observe the entire sky up to declination +55$^\circ$ (Pollacco et al., 2006 ). Each instrument consists of an array of eight 2Kx2K CCD cameras with 200mm f/1.8 photographic lenses.  The pixel scale is approximately 13.7 arc-seconds per pixel, which limits useful observations to regions of the sky away from the galactic plane. Observations are obtained using a broad-band filter over the wavelength range 400-700nm. The instruments are fully robotic and have been fully operational since mid-2006. Observations at defined locations are obtained each night with a cadence of about 5 minutes using a pair of 30-second exposures, resulting in 500 $-$  1000 images being obtained per camera on a clear night. Images are processed to produce aperture photometry at the locations of catalogued stars using a synthetic aperture radius of 48 arc-seconds. A variety of de-trending and transit-detection algorithms are used to identify candidates from the resulting lightcurves (Collier Cameron et al. 2006).

The WASP database currently contains over 2000 nights of data for more than 30 million stars.  This amounts to more than 400 billion photometric measurements that have been measured from almost 1.25 million images. The average number of photometric measurements per star is approximately 13,000, but in some regions of the sky there are 20-30,000 observations per star.

The majority of the follow-up effort in the Southern hemisphere is conducted in collaboration with Geneva Observatory using the Swiss Euler 1.2-m telescope, using both the Coralie spectrograph and the EulerCam CCD camera. High precision transit photometry is also obtained for many targets by Michael Gillon using the Trappist telescope. The HARPS spectrograph on the ESO 3.6-m telescope is also used for confirmation of low-mass planets, particularly for fainter targets. In the Northern hemisphere the majority of exoplanet discoveries are confirmed using the SOPHIE spectrograph on the 1.93-m telescope at the Observatoire Haute Provence following extensive characterisation of candidates using a variety of telescopes and instruments, e.g., the FIES spectrograph on the NOT telescope, the RISE camera on the LT telescope, etc.

To-date the WASP survey has published 67 exoplanet discoveries, including two exoplanets also discovered independently by the HAT survey and a brown dwarf in an eclipsing binary with a mass of 61 Jupiter masses (WASP-30b). Highlights of the WASP survey include the discovery of several hot Jupiter planets with periods of less than one day (WASP-19b, WASP-18b and WASP-43b), the discovery of the first hot Jupiter with a retrograde orbit (WASP-17b) and a hot Jupiter orbiting a pulsating A-type star (WASP-33b). Papers announcing the discovery of approximately 25 further exoplanets are in preparation. The announcement of the 100th planet discovered by the WASP survey is expected within the next year. The total number of planets that are, in principle, discoverable by surveys like WASP is approximately 300, but the practicalities of planet discovery probably limit the actual number that can be discovered by WASP to about half this number.

The exoplanets discovered by the WASP survey typically have K-band magnitudes in the range K = 8 $-$ 11 and orbital periods P = 1 $-$ 6 days. The detection efficiency of the survey drops rapidly for orbital periods longer than about 10 days. The host stars of the WASP exoplanets are typically late F-type stars and early G-type stars, with some K-type host stars. The exoplanets have masses from about 1/3 of a Jupiter mass upwards, with most having masses close to 1 Jupiter mass. Continual improvements to the instrumentation and data processing and the increasing number of observations per star have enabled the survey to discover exoplanets with radii as low as 0.6 Jupiter radii, but the majority of exoplanets discovered have radii of 1 $-$ 1.4 Jupiter radii. 

\begin{figure}
\includegraphics[scale=0.7]{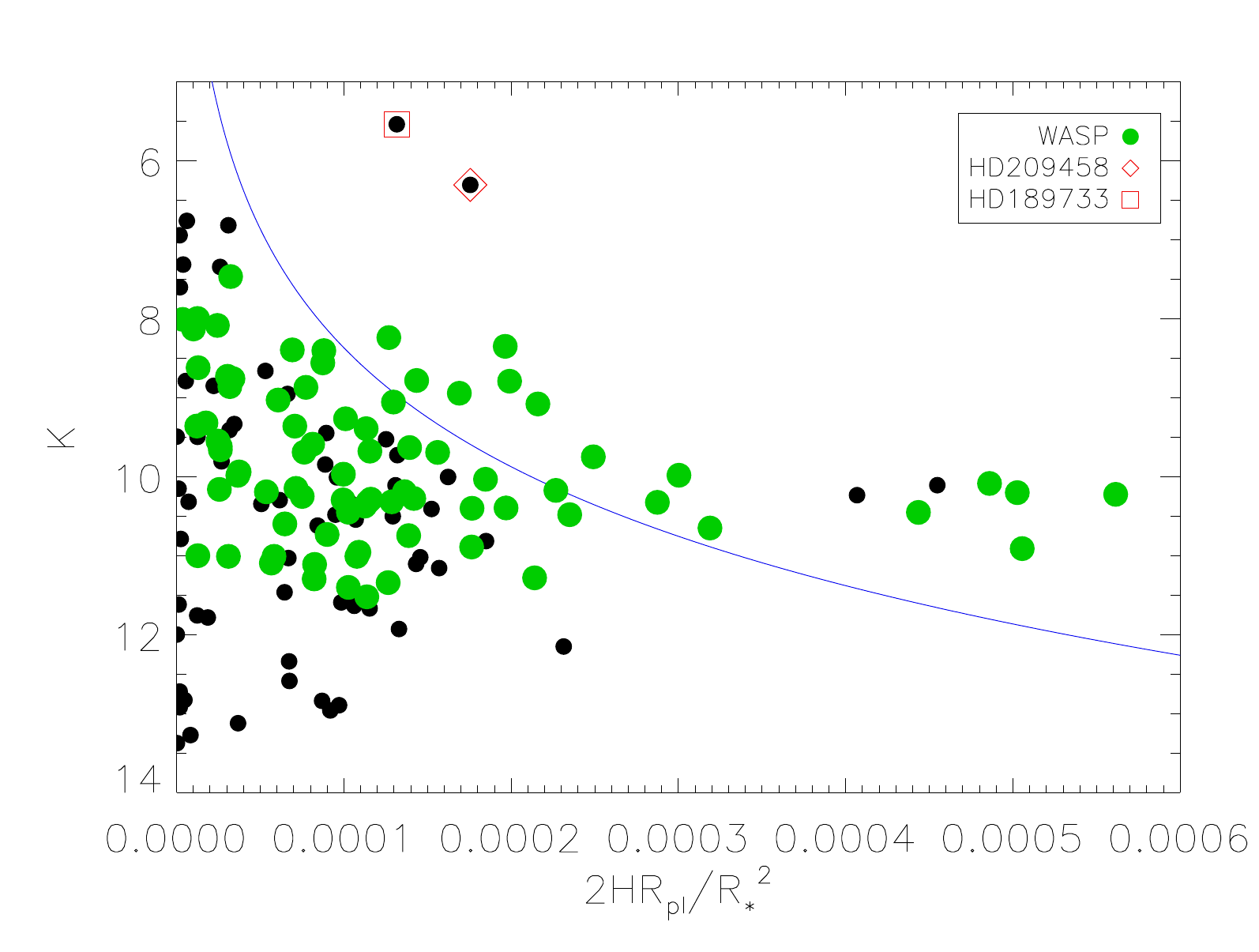}
\caption{Nominal signal strength for transmission spectroscopy of known exoplanets calculated from the atmospheric scale height  (H), the radius of the planet (Rpl) and the radius of the star (R*). The data for some of the WASP exoplanets plotted here are based on very preliminary analyses and so may be subject to large uncertainties. A line of constant signal-to-noise (continuous line) shows that there are several exoplanets that are good targets for transmission spectroscopy that orbit quite faint stars (K $\approx$ 10).}
\label{Fig1}
\end{figure}

The depth of the eclipse produced by the occultation of an exoplanet by its host star can be estimated based on the observed transit depth and an estimate of the equilibrium temperature of the day-side of the exoplanet. For WASP exoplanets, the resulting distribution of eclipse depths is typically 1 $-$ 3.5 parts-per-thousand. Detailed observations of these eclipses using quite short observing sequences (6 $-$ 8 hours) are well within the capabilities of EChO. WASP exoplanets are distributed fairly evenly across the sky and so observations of the secondary eclipses of WASP exoplanets will be valuable scientific observations that can be easily accommodated into the observing schedule. There are many hot Jupiters that have been discovered by the WASP survey that have large radii and low masses, and that orbit bright stars. A nominal value for the strength of the transmission spectroscopy signal can be calculated from the area of the annulus with a width of one atmospheric scale height relative to the size of the star.  These values are shown together with the host star K-band magnitude in Figure \ref{Fig1}. Although there are fewer suitable targets for transmission spectroscopy than for eclipse depth observations, transmission spectroscopy of WASP exoplanets will also be a valuable science goal for EChO that can achieved using observations that make good use of shorter observing windows in the schedule. 

\section{Next Generation Transit Survey }

The Next Generation Transit Survey (NGTS, Chazelas et al. 2012, Wheatley et al. 2013) is a new ground-based transit survey that is being constructed during 2014 at the ESO Paranal Observatory in Chile. The primary objectives are: 1) to identify the first statistically-significant sample of transiting Neptunes and Super-Earths orbiting stars that are sufficiently bright for radial velocity confirmation and mass determination; 2) to identify individual transiting Neptunes and Super-Earths with host stars that are sufficiently bright for secondary eclipse and transmission spectroscopy follow up with JWST, E-ELT and EChO.

The NGTS instrument consists of an array of twelve 20cm f/2.8 telescopes, each mounted on an independently-steerable equatorial fork mount and fitted with a red-sensitive 4Mpix back-illuminated deep-depletion CCD camera. The pixel scale is 5 arcsec, allowing observations relatively close to the Galactic plane. The full instantaneous field of view of the facility is 96 square degrees. Filters limit the bandpass to 550-900nm (with peak sensitivity at 800nm) providing good sensitivity to K and early M dwarfs. 

NGTS builds on heritage from the WASP survey and has been designed to achieve higher photometric precision and hence find smaller planets than has so far been achieved from the ground. The independent fork mounts allow for precise autoguiding and the smaller pixel scale reduces blending and sky noise. Noise models suggest that photometric precision for bright stars (I $<$ 10) will be limited on a 1 hour timescale to 0.2 mmag by scintillation, with faint stars limited by sky noise (I $>$ 14), and intermediate stars limited by source shot noise (14 $>$ I $>$ 10). 

The design has been tested through the operation of a prototype instrument on La Palma during 2009-2010, with 0.3mmag precision demonstrated in ensemble photometry over a wide field. Testing of a complete NGTS telescope unit has also demonstrated sub-mmag photometry from Geneva, Switzerland. The full facility will benefit from the exceptional photometric conditions of the ESO Paranal site, as well as the excellent follow up capabilities of co-located ESO facilities such as HARPS, the VLT and the E-ELT.

The baseline plan is for a five-year science survey running from 2014 to 2019, with four fields covered intensively each year. The full NGTS survey will therefore cover a sky area of about 1920 square degrees, which is eighteen times the coverage of Kepler. This survey area will be spread approximately evenly in right ascension, and concentrated in a declination range between -30 and -50 degrees. Simulations of the survey window function based on real weather statistics of the Paranal site suggest more than 50\% detection efficiency for planetary orbital periods up to 16 days. 

Simulations of the NGTS planet catch have been made based on populations of host stars drawn from the Besan\c{c}on model of the Galaxy (Robin et al. 2003) with stellar spectra from the Pickles atlas (Pickles 1998) and representative sky spectra measured on La Palma. These simulated host stars were randomly assigned a population of planets based on the planet size distributions and occurrence rates from Kepler (Howard et al. 2012). The detectability of each simulated planet was then assessed using the noise model for the NGTS facility, which is based on the known photometric performance of the prototype instruments. 

\begin{figure}
\includegraphics[scale=0.7]{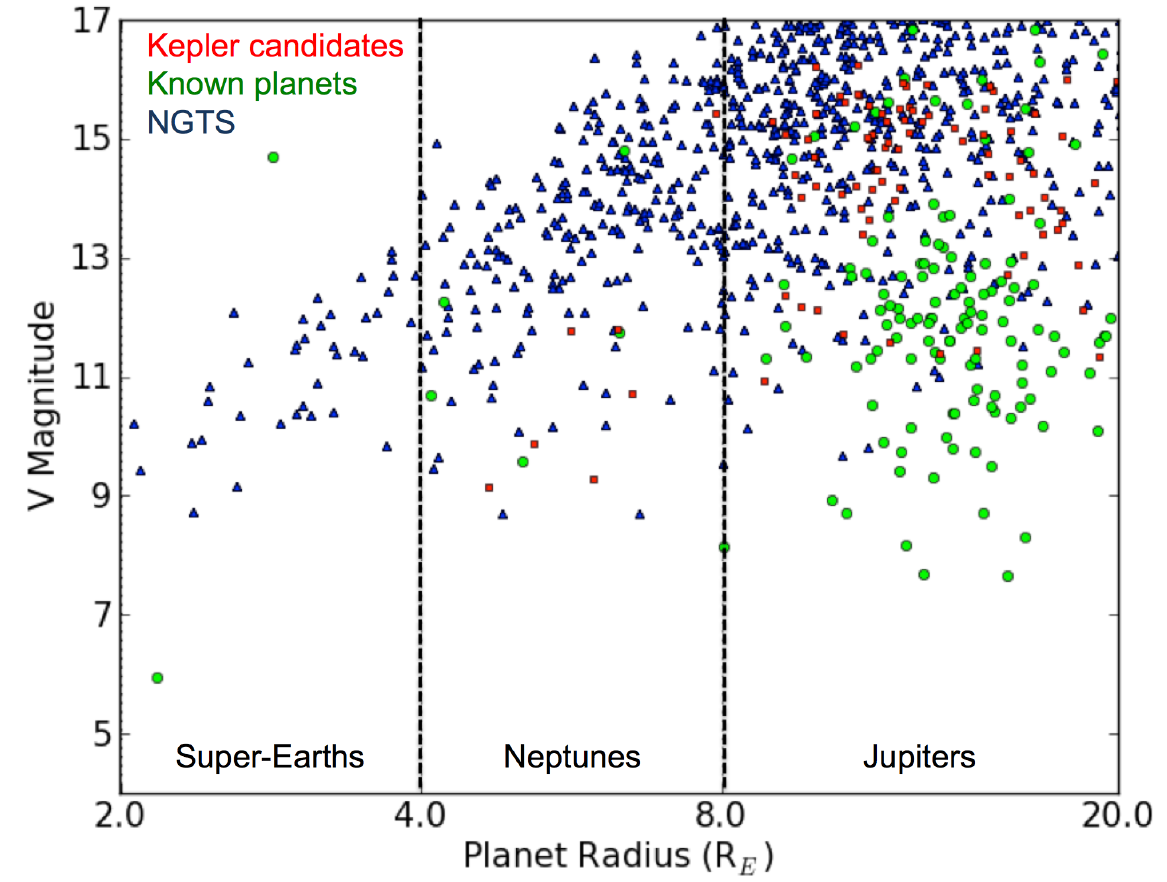}
\caption{The simulated confirmable planet population from NGTS. This assumes a survey of 1920 square degrees over five years. Each of the plotted simulated planets can be confirmed with HARPS or ESPRESSO in less than 10 h exposure time. This instance of the simulation shows 39 confirmable super-Earths and 231 Neptunes. Of these, 23 super-Earths and 25 Neptunes orbit stars brighter than I=11}\label{Fig2}
\end{figure}

The resulting NGTS planet candidate population has been filtered with the sensitivity limits of the HARPS and ESPRESSO radial-velocity spectrographs, and the final predicted confirmable population is shown in Figure \ref{Fig2}. Assuming a time allocation for each candidate of 10 hours with HARPS or ESPRESSO, the predicted confirmable planet population for NGTS includes 231 Neptunes and 39 super-Earths (with these totals obviously subject to counting statistics). 

Of the simulated confirmable planets, 23 super-Earths and 25 Neptunes orbit stars brighter than I=11, and are likely to be suitable for detailed atmospheric characterisation. Of these planets, most orbit G and K dwarf stars, with orbital periods in the range 4 to 12 days.

The high photometric precision of NGTS will also allow Jupiter sized planets to be identified from a single transit, extending the sensitivity of the survey to warm Jupiters and probably a few in the habitable zone. The orbital periods of these systems will be determined with radial velocity follow up, and their atmospheric characterisation with EChO will explore the sensitivity of atmospheric chemistry to stellar insolation. 

\section{HATNet and HATSouth}
\label{sec4}

\subsubsection*{HATNet} 
HATNet is 2-station (Arizona, Hawaii) network of automated, small and wide field telescopes (Bakos et al. 2002; 2004). HATNet utilizes 6 identical instruments, each with an 11 cm aperture f/1.8 lens and a 4K x 4K front-illuminated CCD with 9ÕÕ pixels (yielding a 10.6$^\circ \times 10.6^\circ$  field). Each instrument, called a HAT, can attain a photometric precision reaching 4 mmag at 3.5-min cadence on the bright end at r $\sim$ 9.5, and 10 mmag at r $\sim$ 12.1. HATNet has been operational for 9 years, the stations opening up on 2000 nights yielding some 1.6 million science frames, covering ~19\% of the sky. Each selected area has been monitored for $\sim$ 2 months. The simultaneous sky coverage of HATNet is approximately 400 sq degrees, since telescopes share fields.

HATNet has been a major contributor to the discovery of transiting extrasolar planets (TEPs). The $\sim$ 50 HAT planets constitute roughly one fourth of the known population with accurate mass determinations. These planets exhibit an amazing diversity, with masses ranging from that of Neptune (0.08M$_J$; HAT-P-11b) to 9 times the mass of Jupiter (9M$_J$; HAT-P-2b), and radii from half of Jupiter to twice that of Jupiter. HATNet is the only ground based survey that found Neptune size/mass transiting exoplanets (HAT-P-11b and HAT-P-26b).

Expected Yields:  With HATNet we currently achieve better than 1\% photometry at 3.5 minute cadence for stars in the interval 9.4 $<$ r $<$ 12.1, and better than 2\% for stars with 7 $<$  r  $<$  13.2. The transit depths we can detect with such light curves are, of course, much smaller than 1\% or 2\%. For reference, a 2.3 d period, 5 mmag deep transit can be detected with an S/N of ~$\sim$ 8 for a typical light curve with 10,000 points, 1\% precision and typical red noise. Using the Besan\c{c}on Galactic model (Robin et al. 2003), we estimate that we obtain 1\% photometry with HATNet for 25 F2--M9 dwarf stars per square degree and 2\% photometry for 75 dwarfs per square degree when observing a field at a typical Galactic longitude and latitude of l = 45$^\circ$, b = 30$^\circ$. In the 12 selected fields we monitor per year, we thus observe a total of $\sim$30,000 dwarfs with better than 1\% photometry, and ~90,000 dwarfs with better than 2\% photometry. Of particular interest are lower mass stars whose smaller radii facilitate the detection of smaller planets. A photometric precision of 1\% per 3.5 minutes is sufficient to find transiting Neptune-size planets around K0--K9 dwarfs (as demonstrated, e.g., by HAT-P-11 and HAT-P-26). We estimate that $\sim$  4000 K0ÐK9 dwarfs are monitored per year with HATNet at better than 1\% precision. Even smaller planets, such as super-Earths, could be detected around M3--M9 dwarfs (e.g. GJ 1214; Charbonneau et al. 2009). We estimate that $\sim$ 50 M3--M9 dwarfs are monitored per year with HATNet at better than 1\% precision, and 170 per year at better than 2\% precision.

We currently identify $\sim$ 300 candidates per year, and after the confirmation procedure we conclude that we find 1 TEP per every  $\sim$  4200 dwarf stars surveyed with better than 1\% photometry and 1 TEP per every $\sim$ 20000 dwarf stars surveyed with 1Ð2\% photometry. Based on the current candidate identification rate, pending adequate follow-up resources, we then expect to discover $\sim$ 10 TEPs per year with HATNet.

Regarding the stellar metallicity, age and temperature parameter space, we observe a magnitude limited sample. The majority of these dwarfs are F and G spectral type. Regarding the planet temperature, mass and density, we are mostly sensitive to Saturn and Jupiter sized planets, with a few expected Neptune detections per year. Radial velocity follow-up tools we have been routinely using ensure confirmation of Neptune-mass objects.

\subsubsection*{HATSouth}

HATSouth (Bakos et al. 2013) is the Southern hemisphere network counterpart to HATNet. We made several key changes with respect to HATNet to this network, most notably placing instruments at three locations in the Southern hemisphere (Chile, Namibia, Australia), which provide near round-the-clock monitoring of 128 square degrees at any time. Two instruments are located at each of the three HAT-South sites, and each instrument consists of four 18 cm aperture f/2.8 Takahashi hyperbolic astrographs imaging onto separate Apogee U16m 4K x 4K CCD detectors. The mosaic field of each instrument is 8.2$^\circ$ x  8.2$^\circ$  imaged onto 8K x 8K pixels with a scale of 3.7'' pixel$^{-1}$. The observations are carried out through Sloan r band, similarly to HATNet. HATSouth has been fully operational for three years, instruments have gathered 1.9 million science frames covering 5\% of the sky. HATSouth has discovered $\sim$  300 TEP candidates, primarily hot Jupiters and Saturns. Three of these have been confirmed and published (Penev et al. 2013, e.g.), and another dozen are in their final stages of confirmation.

Expected yields: In order to estimate the expected yield of transiting planets from the HATSouth survey we conducted transit injection and recovery simulations, as described in Bakos et al. (2013). These simulations use the Besan\c{c}on Galactic model (Robin et al. 2003), planet occurrence rates from the Kepler mission (Howard et al. 2012), realistic noise time-sampling (using existing HATSouth light curves).

Using the Besan\c{c}on Galactic model, we estimate that we obtain 1\% photometry with HATSouth for 137 F2--M9 dwarf stars per square degree and 2\% photometry for 370 dwarfs per square degree when observing a field at a typical Galactic longitude and latitude. In the 12 selected fields we monitor per year, we thus observe a total of $\sim$ 100,000 dwarfs with better than 1\% photometry, and $\sim$ 290,000 dwarfs with better than 2\% photometry. We estimate that $\sim$  12,000 K0--K9 dwarfs are monitored per year with HATSouth at better than 1\% precision, and $\sim$  770 M3 to M9 dwarfs at better than 2\% precision.

Assuming 12 fields observed per year, HATSouth is capable of finding $\sim$  30 transiting planets per year (as mentioned before, pending availability of follow-up resources), including $\sim$ 1 planet per year with R $<$  0.7RJ and $\sim$  6 planets per year with P $>$10 d.

Due to the bigger optics, HATSouth reaches to fainter magnitudes than HATNet, and has more late dwarf stars in it sample. Due to the networked setup and round-the-clock monitoring, we reach out to longer periods and shallower transits. HATSouth is more efficient in the detection of Neptunes. HAT and HATSouth exoplanets are distributed homogeneously across the sky, simplifying scheduling observations with EChO.

\section{GAIA}
\label{sec5}

Gaia (e.g., Lindegren 2010), launched in December 2013, in its five-year all-sky survey will deliver high-precision global astrometry and complementary spectrophotometry and spectroscopy for nearby main-sequence stars down to $\sim$ 20. 

Gaia will be capable of measuring accurate astrometric orbits and masses for giant planets within approximately 0.3-4 AU of moderately bright (6$<$V$<$13, with current attempts at pushing the bright limit down to V$\sim$ 2-3) F-G-K stars out to $\sim$ 200 pc from the Sun (Casertano et al. 2008). Depending on the actual bright magnitude limit (and corresponding astrometric precision) ultimately set for Gaia, approximately 30\% of presently known RV planets might be seen in Gaia astrometry (assuming their true masses do not differ significantly from their minimum mass estimates). A volume-limited (D$<$30 pc) sample of thousands of low-mass M dwarfs down to the V=20 mag limit of the survey will also be screened for 'cold' giant planets (Sozzetti et al. 2014) with a similar separation range, while $10^4$ bright (J$<$10) M stars with spectral sub-type earlier than M5 within approximately 60 pc from the Sun (Lepine \& Gaidos 2011) will also be observed by Gaia with enough astrometric sensitivity for giant planet detection.
 
\begin{figure}
\includegraphics[]{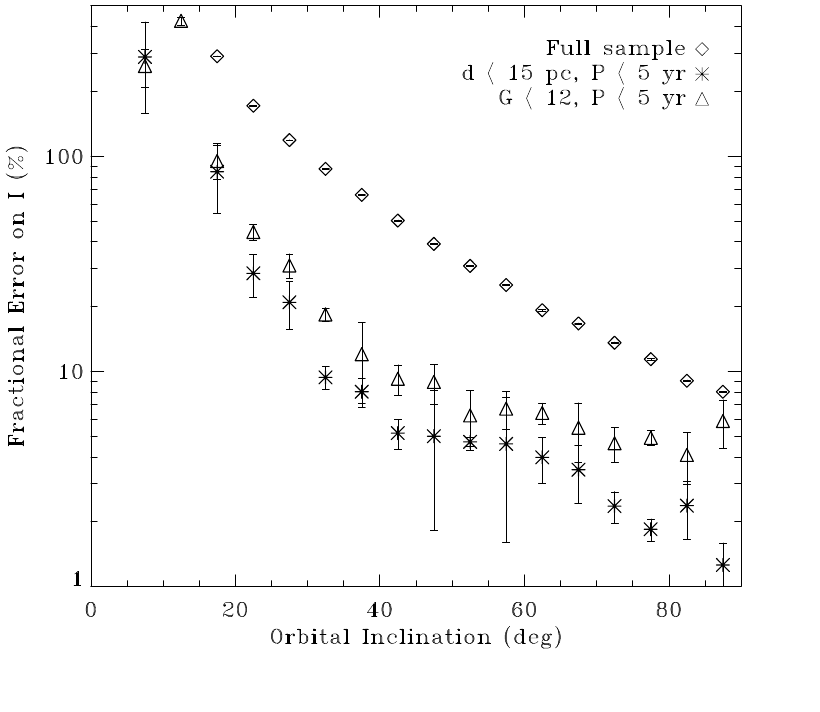}
\caption{Fractional error on the inclination angle as a function of the angle itself as determined from Keplerian orbital solutions for Jupiter-mass companions orbiting a sample of $\sim$ 3000 known M dwarfs within 33 pc from the Sun in the simulations of Sozzetti et al. (2014). Well-sampled orbits close to edge-on configurations might have the inclination angle (i) measured with uncertainties of a few degrees.}\label{fig.gaia}
\end{figure}

Well-determined orbits will allow to measure accurately the orbital inclination (within a precision of a few degrees, Figure \ref{fig.gaia}). Up to 10,000 NEW giant planets might be found with Gaia, orbiting stars of all spectral types, ages, and metallicities. Based on simple geometric considerations, it is thus possible that cold Jupiters orbiting stellar hosts with a very wide range of properties (mass, metallicity, age) might be found in transiting configurations. Extrapolations to the reservoir of low-mass M dwarfs within 100 pc from the Sun (Sozzetti et al. 2014) indicate that a few tens of such systems could indeed be identified. Even before EChO launches, such objects might have been confirmed to be transiting planets via follow-up photometric measurements, and thus would help to populate the EChO target list with new systems in a separation range (typically, 0.5-3 AU) completely inaccessible by present-day ground-based transit surveys, and even only marginally probed by space-borne instruments such as Kepler. However, on the one hand, cold giant planets detected by Gaia (both new and already known from RV work) will usually NOT be transiting. On the other hand, when the full orbit and actual companion mass are available and measured with good accuracy, it is then possible to predict where and when one will find the planet around the star. In particular, for the purpose of detection in combined light with EChO, it will be possible to predict the epoch and location of maximum brightness. For non-transiting eccentric systems, this will be extremely valuable information as it might be possible to study the seasonal changes in atmospheric composition due to the significantly variable irradiation conditions (for an e=0.6 orbit, the stellar flux varies by a factor of 16 along the planet's orbit!). As a result, a large reservoir of potential giant planet hosting target systems will then become available to EChO, from which it will be possible to select a sample of high-priority systems, based on their actual orbital architectures, masses and primary properties.
 
\section{APACHE and HARPS-N}

HARPS-N is an \'echelle spectrograph that covers the visible wavelength range between 383 and 693 nm (Cosentino et al. 2012). It is a near-twin of the HARPS instrument mounted at the ESO 3.6-m telescope in La Silla. It was installed at the TNG in March 2012. After instrument commissioning in mid 2012, it was offered for open-time programmes starting in August 2012. 

The large programme Global Architecture of Planetary Systems (GAPS) is using, on a competitive basis, the unique capabilities provided by HARPS-N to maximize the scientific return in several aspects of exoplanetary astrophysics. The GAPS programme is composed of three main elements, including a) radial-velocity searches for low-mass planets around stars with and without known planets over a broad range of properties (mass, metallicity) of the hosts, b) characterization measurements of known transiting systems, and c) improved determinations of relevant physical parameters (masses, radii, ages) and of the degree of star-planet interactions for selected planet hosts. 

\begin{figure*}
\includegraphics[]{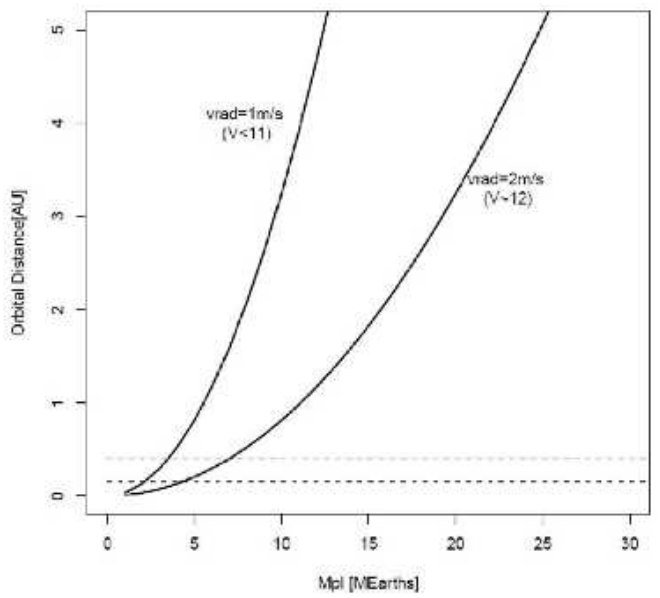}
\includegraphics[]{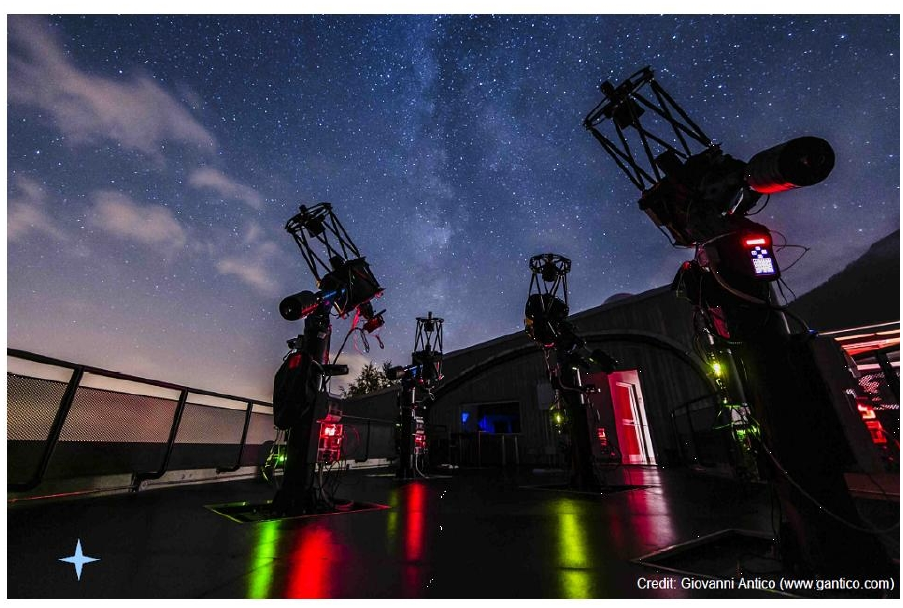}
\caption{: Left: Maximum orbital distance as a function of planetary mass achievable with HARPS-N observations for a 0.5 M$_E$ star (for two regimes of single-measurement error and target magnitude). Right: Close-up night view of the APACHE telescope array at the OAVdA site.}\label{fig.apache}
\end{figure*}

One of the GAPS program elements focuses on  HARPS-N search for potentially habitable Neptunes and Super-Earths (M $<$ 15 M$_{E}$) around a well-defined, well-characterized sample of early M0-M2 dwarf stars. The HARPS-N observations will allow us to a) achieve uniform sensitivity to habitable-zone Super Earths around M dwarfs, thus vastly improving on the precision of early estimates of their occurrence rate $\eta$, b) characterize the dependence of planetary systems architecture on stellar mass, and c) possibly provide benchmark transiting systems suitable for follow-up observations aiming at atmospheric characterization with EChO. For this reason this sample is monitored photometrically by the APACHE transiting planet search program. Based on HARPS-S data (Bonfils et al. 2013) we expect that the GAPS project will detect  $\sim$ 30 Super Earths or mini Neptunes with orbital period P $<$ 100d (see Figure \ref{fig.apache}).

APACHE (A Pathway to the Characterization of Habitable Earths) is the first Europe-based photometric transit search for small-size planets orbiting $\sim$ 3000 bright, low-mass M dwarfs (Sozzetti et al. 2013). The APACHE survey officially started in July 2012 at the site of the Astronomical Observatory of the Autonomous Region of the Aosta Valley (OAVdA), in the Western Italian Alps. APACHE employs an array of five 40-cm class telescopes hosted on a single platform with electronically controlled roll-off enclosure (Fig. \ref{fig.apache}). The arrangement of the array has been defined so as to maximize sky coverage during campaign operations. The telescope array is composed of five identical Carbon Truss 40-cm f/8.4 Ritchey-Chr\'etien telescopes, with a GM2000 10-MICRON mount and equipped with a FLI Proline PL1001E-2 CCD Camera and Johnson-Cousins R \& I filters. These systems guarantee state-of-the-art quality performance: negligible temperature gradients, QE 80\% in the whole wavelength range of interest. The open source observatory manager RTS2 (Kubanek 2010) is the choice for the high-level software control of the five-telescope system, including dynamic scheduling of the observations. The APACHE Input Catalogue (AIC) of M dwarfs is constructed starting from the all-sky sample of 8889 bright (J $<$ 10) low-mass stars in L\'epine \& Gaidos (2011).

APACHE draws on the pioneering work carried out by the MEarth project (Nutzman \& Charbonneau 2008), in that is adopts a one target per field approach and the corresponding constraints on the observing strategy optimized for the array of five APACHE telescopes and the chosen site characteristics. It extends and complements MEarth in the Northern hemisphere, in that it targets early- to mid-M dwarfs, conceding in detectable planet radius for fixed photometric precision but focusing by construction on a target sample in principle more favorable for atmospheric characterization work in case of positive detection. Now well into the second year of operations, the program is delivering the expected quality in differential photometric precision (median $\sim$ 5 mmag, sufficient for a 3-$\sigma$ detection of transiting exoplanets with 2-4 Earth radii on short-period orbits), and the first short-period transit candidates are being vetted. The synergy potential between APACHE, Doppler (e.g., HARPS-N, near-infrared spectrographs such as GIANO), astrometric (Gaia), and atmospheric characterization (e.g., EChO) programs is potentially huge. For example, transiting exoplanets discovered by APACHE with precise mass (from HARPS-N) and radius (from Gaia) determinations will orbit by construction M dwarfs bright enough to enable detailed characterization of their planetary interior structure and atmospheres with space-borne observatories such as EChO. Extrapolating the results of Dressing  \& Charbonneau (2013), APACHE will detect about half dozen of planets with radius in the 1.4-4.0 REarth range and periods $<$ 10 days.

\section{TESS}

The Transiting Exoplanet Survey Satellite is a NASA Explorer Program mission (Ricker et al. 2010, space.mit.edu$/$TESS$/$TESS$/$TESS$\_$Overview.html)  selected for launch in 2017. TESS will use four wide-field imaging CCD cameras, each with a diameter of $\sim$ 9.6 cm, and a field of view of about $23^\circ \times 23^\circ$. The mission will last two years and will monitor $\sim$ 45,000 sq. degrees of sky (an area about 400 times larger than the one covered by the Kepler Mission). The main targets will be bright nearby G-K stars with apparent magnitudes brighter than about I = 12.0 $-$ 13.0 and M-dwarfs brighter than I $\sim$ 13.

\begin{figure*}
\includegraphics[scale=0.6]{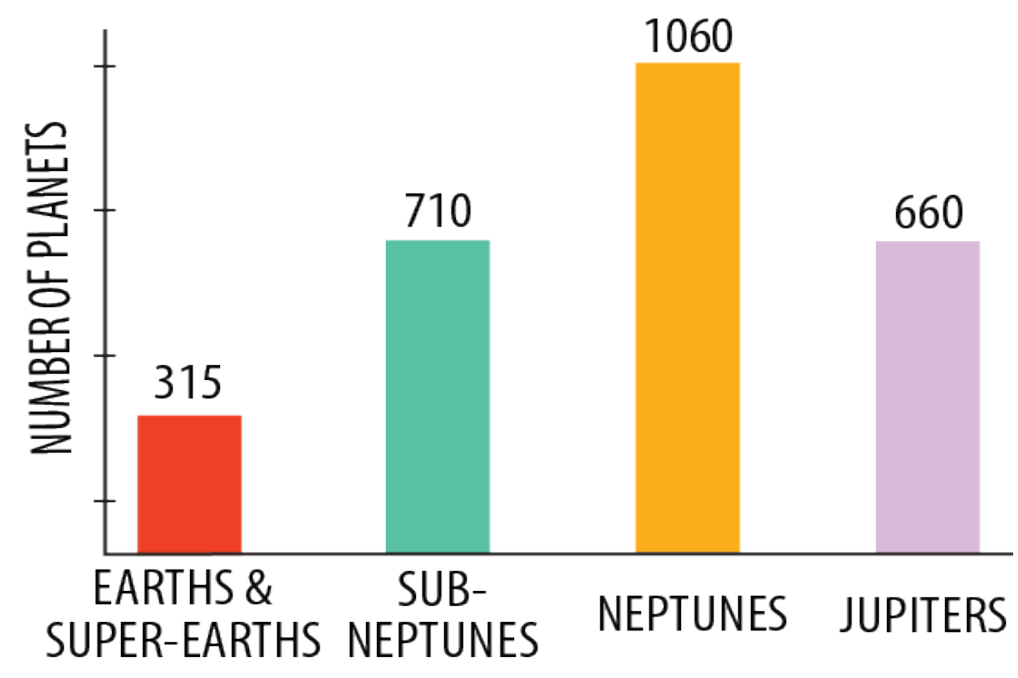}
\includegraphics[scale=0.6]{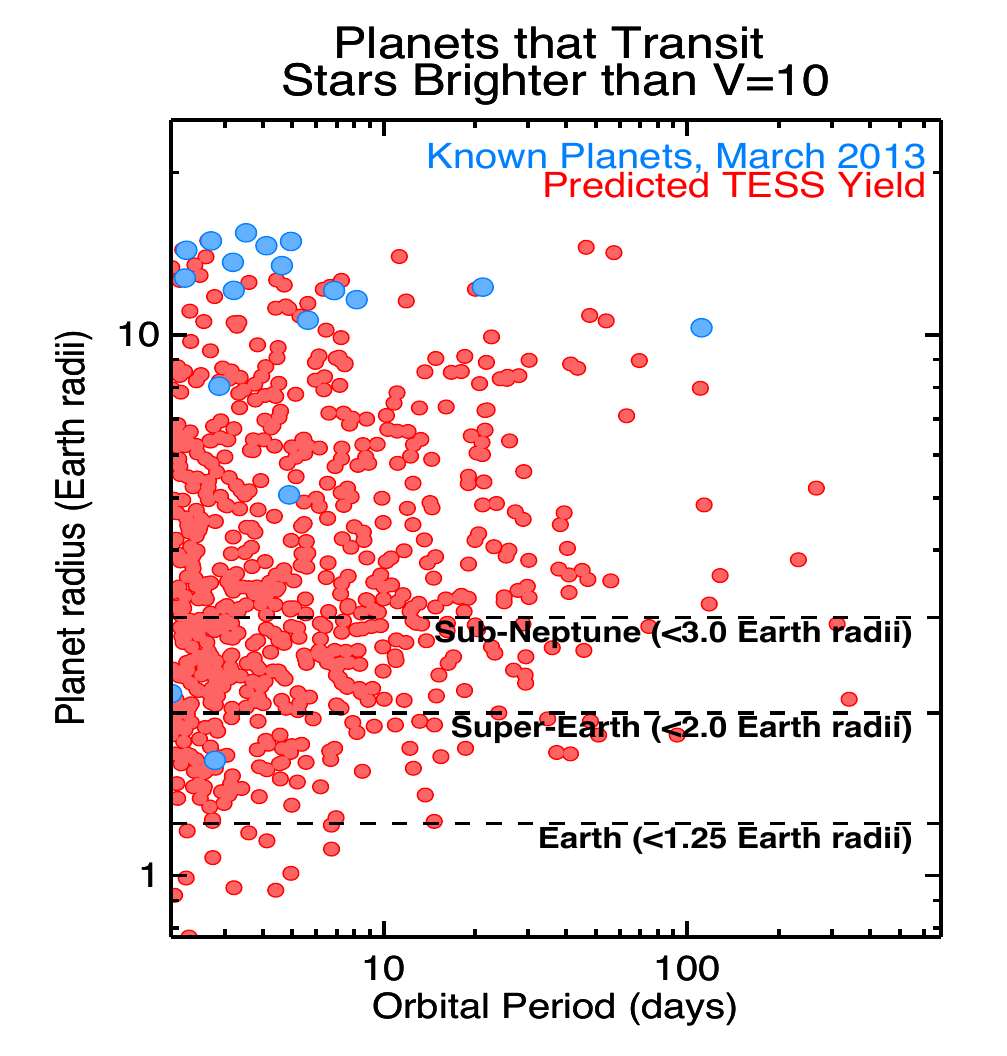}
\caption{: Left: Expected science yield from the TESS mission. Right: : Radius-Orbital period distribution of transiting exoplanets found around nearby stars brighter than V=10 as until March 2013 (blue dots), versus the number of such planets expected to be discovered by TESS (red dots).}\label{fig.tess}
\end{figure*}

Upon completion of the mission in 2019-2020, TESS will have observed about $2.5 \times  10^6$ stars. Of those, $\sim$ 5000 objects are expected to reveal transit-like signals and therefore be classified as planet candidates. After candidate validation, the predicted science yield of TESS is about 300 Earths and Super-Earths, as illustrated in Figure \ref{fig.tess}, in addition to about 2,400 planets in the radius range between Jupiters and sub-Neptunes (Rp $<$ 2.0 R$_E$) (Seager et al. 2010). The majority of TESSÕ candidates will have been validated before EChO's launch.

Given the TESS observing strategy, all the planets will be orbiting stars brighter than V $\sim$ 12 and the majority will have orbital periods shorter than 10 days , with some up to 100 days. Many of them will be suitable for EChO follow-up, as shown for example in Figure \ref{fig.tess}, where we represent all the transiting planets known as of March 2013 around stars brighter than V=10, versus the expected number of planets around stars of similar magnitude once the TESS mission is complete.

\begin{figure}
\includegraphics[scale=0.6]{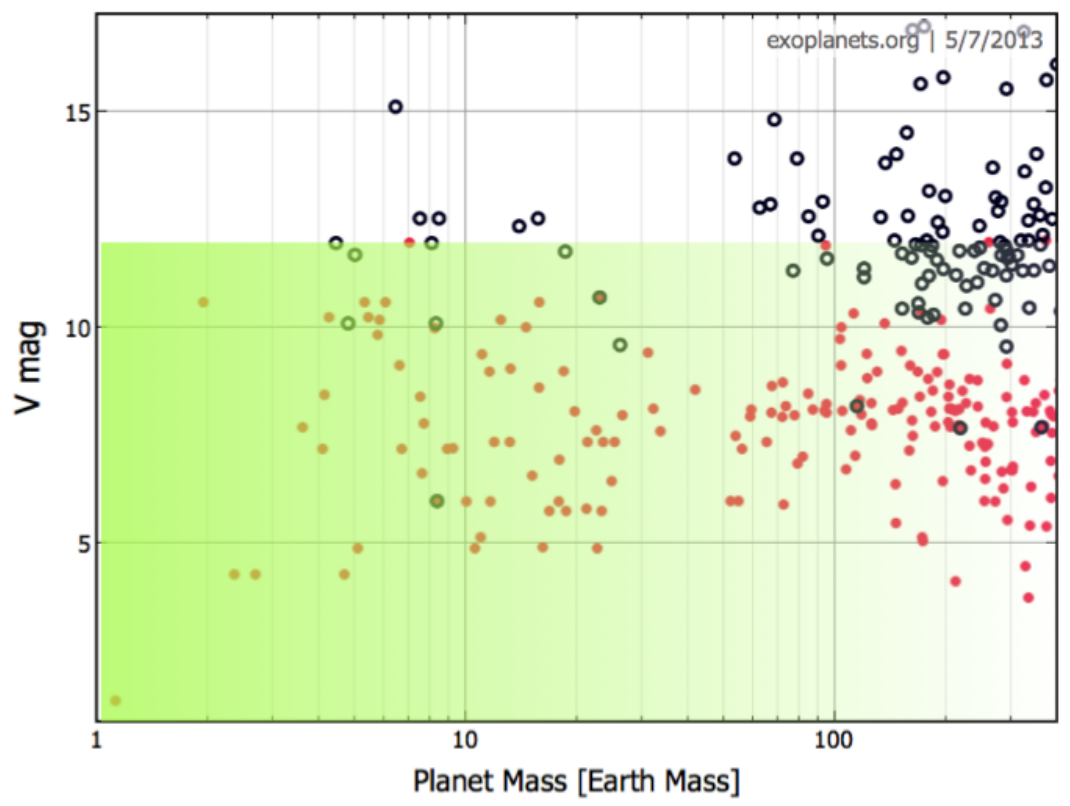}
\caption{Planets with measured mass from RV survey (red dots). Planets with measured radius from transit survey (black circles). The green shaded area is where CHEOPS will provide accurate radius measurements.}\label{fig.cheops}
\end{figure}

\section{CHEOPS}

The CHaracterizing ExoPlanet Satellite (CHEOPS) has been selected by ESA as its first Small mission in the framework of the Cosmic Vision program to search for transits on bright stars already known to host planets in the super-Earth to Neptune mass range (1 $<$ Mplanet/MEarth $<$  20). The launch is planned at the end of 2017, for an operation life of 3.5 yrs. 

The main science goal of the CHEOPS mission (Broeg et al. 2013) will be measuring the radius of planets transiting bright stars to 10\% accuracy with the aim to study the structure of exoplanets smaller than Saturn orbiting bright stars. With an accurate knowledge of masses and radii for an unprecedented sample of planets, CHEOPS will set new constraints on the structure and hence on the formation and evolution of planets in this mass range and on habitability (Alibert 2014), and will provide unique targets for future ground (e.g., E-ELT) and space-based (e.g., JWST, EChO) facilities with spectroscopic capabilities. 

Large ground-based high-precision Doppler spectroscopic surveys carried out during the last years have identified nearly a hundred stars hosting planets in the super-Earth to Neptune mass range (1 $<$ Mplanet/MEarth $<$ 20, Mayor et al. 2011). As search programs continue, the number is going to increase in the coming years. On the other hand, only few of these planets have accurate measurements of the radius (cf. Figure \ref{fig.cheops}).

CHEOPS payload is made by a F/8 Ritchey-Chretien style on-axis telescope with $\sim$ 30cm effective aperture providing a de-focussed image of the target star in the visible (400 to 1100 nm) with a flat point spread function (PSF) devoid of structures at high spatial frequencies. The detector will be a single frame-transfer back-side illuminated CCD detector, thermally stabilized to 5Ð15 mK at an operating temperature of -40 C. To obtain high photometric stability, thermal stability of the instrument and straylight suppression from the Earth are design drivers.

CHEOPS main products are ultrahigh precision photometric series, centred on the time of the expected transits (from the known planetary ephemerids). When a transit is detected, the planetary radius Ð relative to the stellar radius Ð can be measured. The joined information on planetary mass (from RV data, already known for CHEOPS primary targets) and radius (from the transits) will allow CHEOPS to reach its main goal, i.e. the determination of the mass-radius relation in the sub-Neptune mass regime. The determination of the bulk density of the planets, provides direct insights into the composition of the body and enables a quantitative constraint on the upper limit on the planetÕs envelope mass. For instance, the presence of a gaseous envelope (only a few percents in mass) or icy mantle (above 10\% in mass) has a large effect on the planet radius and mean density; hence CHEOPS will be able to discriminate between telluric, and hydrogen-rich or ocean-planets, and will identify planets with significant atmospheres as a function of their mass, distance to the star, and stellar parameters, that are of interest to EChO. 

CHEOPS will also measure the phase modulation due to the different contribution of the dayside of hot Jupiter planets and in some cases it will measure the secondary eclipse (see e.g. Demory et al. 2013) . These measurements provide information about the energy flux in the atmosphere of the planet. EChO can use such CHEOPS measurements to define/refine the sample worthy spectroscopic investigation.

CHEOPS will be orbiting in a Sun Synchronous Low Earth Orbit (LEO) having a local time of ascending node (LTAN) of 6 am and an altitude in the range of 620 to 800 km (depending on launch opportunities). Hence, the satellite will follow as close as possible the day-night terminator and the target stars will be above the night side of the Earth. CHEOPS will be able to point at nearly any location on the sky; for each target it will be possible to have till 50 days of consecutive observations per year, with a temporal resolution of 1 minute. 

The CHEOPS input catalogue is under construction and criteria for target selection are currently studied. 

As general principle, CHEOPS has two main kind of targets: i) $\sim$150-200 (very) bright stars (6$\leq$ I $\leq$ 9 mag) with a known planet from RV searches, and ii) $\sim$ 50 bright stars (V$\leq$ 12 mag) with a known transit from ground-based transit searches. The first sample is fully accessible to EChO as far as brightness is concerned.

A dozen of transiting planets in the RV sample are expected to be identified. These will be the targets for which accurate planetary structure, and insight on the nature of the atmosphere will be derived. The follow up of ground-based detected transiting planets, will results in accurate measurement of radius for about 22, 22, and 6 planets with radius in the range 1.5 -10 REarth, 1.5-4 REarth, and $<$1.5REarth, respectively.

It is worth to note that the CHEOPS mission operation is conceived in order to adjust the program to accommodate new and interesting objects. E.g. CHEOPS could follow-up any new targets identified by TESS requiring more data, or include in its schedule any new interesting targets whose characterization could be of interest to plan follow up spectroscopic measurements. 

\section{Conclusions}

Previous sections describe a (non-exhaustive) list of surveys that will provide targets for the mission EChO. For each survey the most relevant features for ECHO are highlighted, in particular the stellar magnitude, the mass of the planets and their distance from the star. In  table \ref{tab.predicted} we summarise  what types of planets suitable to be observed by EChO will be provided by these surveys. Table 1 shows that in the next few years we will have available the optimal targets needed to fulfill the EChO objectives.

\begin{table}
\begin{center}
\caption{Transiting planets expected to be discovered in the next decade suitable for EChO observations}\label{tab.predicted}
\begin{tabular}{llllll}
 \hline\hline
Survey&	Stars ÐSpty	&Jupiters	&Neptunes	&Super-Earths	&Notes\\
\hline
WASP/SuperWasp&	G/early-K	&100 	&-&	-	&Porb $<$ 10d\\
NGTS	&G/K/M	&100	&20&	20	\\
HATNet/HATSouth	&K/M	&hundreds	&Tens	&few&	9 years of obs.\\
GAIA	&M&10-15 &      	-	& -	&0.5-3AU\\
HARPS-N/ APACHE&	M&	-&	-	&30/6	&Porb 10-100 days\\
TESS&	G/K/M	&650	&1000 nept +700 sub-nept	&300 (Earth \& S-Earth)&	Porb $<$ 100d\\
CHEOPS	&G/K/M	&$\sim$22	&$\sim$30	&$\sim$18&	\\
\end{tabular}
\end{center}
\end{table}

\subsubsection*{Acknowledgements:}
We acknowledge partial support by the ASI/INAF contract I/022/12/0

\end{document}